# Students Exeat Monitoring System Using Fingerprint Biometric Authentication and Mobile Short Message Service

[1]Olaniyi, O.M,  [2]Omotosho. A, [3]Oluwatosin E.A, [4]Adegoke M.A, [5]Akinmukomi, T.
[1]Department of Computer Engineering
[1]Federal University of Technology, Minna ,Niger-state, Nigeria.
[2,3,4,5]Department of Computer Science and Technology
[2,3,4,5]Bells University of Technology, Ota , Ogun-state, Nigeria.
E-mail: [1]engrolaniyi09@yahoo.com, [2]bayosite2000@yahoo.com ,

**ABSTRACT**
Exeat is a generic term commonly used to describe a period of absence from a centre of learning either for entire day, or parts of a day for appointments, interviews, open days and other fixtures in privately owned academic environment. The current method of monitoring student's movement is inefficient and brings difficulty to the University Halls management checking student's exit/entry into the halls of residence as well as impersonation. By using nexus combination of Ubiquitous Mobile Computing Technology through Mobile Short Message Service and biometric fingerprint approach exeat management and monitoring is quick and easy. Result after testing of the designed and simulated system shows that exeat monitoring systems is less prone to forgery as stakeholders are carried along, capable of preventing impersonation among students, and provide absolute electronic compliance to the policy of issuing exeat to students in the University Halls of Residence.

## 1.0 INTRODUCTION

The increasing use of technology in all aspects of society makes confident, creative and productive use of Information and Communication Technology (ICT) an essential skill for life. ICT capability encompasses not only the mastery of technical skills and techniques, it also facilitates the understanding of these skills in learning, everyday life and employment. ICT capabilities are fundamental to participation and engagement in modern society (The Global Information Technology Report, 2008).

In Information Technology (IT), biometrics refers to technologies for measuring and analyzing human physiological characteristics such as fingerprints, eye retinas and irises, voice patterns, facial patterns, and hand measurements, especially for authentication purposes. Examples of behavioural measureable characteristics include signature recognition, gait recognition, speaker recognition and typing recognition. Biometrics authentication is by measuring a person's physiological or behavioral features. In the past, the common perception of biometrics was that they were limited to use by government facilities and high security areas. However, biometrics is becoming more prevalent in day-to-day applications. It is a type of verification that can be used for authentication when using computers for a variety of purposes.

The concept of exeat is most commonly used to describe a period of absence from a centre of learning. It is also used at certain colleges to define a required note to take absence from school either for entire days, or parts of a day for appointments, interviews, open days and other fixtures (Frischolz  2000). Access control is concerned with determining allowed activities of legitimate users, mediating every attempt by a user to access a resource in the system. Several means and technique have been adopted to restrict access to various domains of human endeavors( Omidiora  2009), but not much has been done with regards to biometrics exeat monitoring system in the privately owned academic domain such as in the Bells University of Technology Ota, Nigeria. The current paper tally approach to exeat has been found to be inadequate because it can be forged or duplicated and does not provide a reliable student monitoring.

The combination of biometric and mobile SMS technology would improve the existing protocol of issuing and managing exeat to students. In literature, different biometric technologies have been applied for verifying users for access to different sensitive places according to the level of security required. These include: Level one Access: ID cards, card keys; Level two Access: PIN, passwords, secret questions.; Level three Access: finger, facial, iris, gait, voice, handprint.; Level four Access: level one + level three. The technologies help in restricting access to the system, allowing access to only those who own a gate card or id card(level 1), know a specific code(level 2), have determined physical mark(level 4), or have a combination  on of both card keys and have a determined physical mark (level 4 mainly for advance systems (Matyas and  Riha 2000) .

This work proposed exeat system on the level three for absolute solution to the required security measure for the problem domain. This proposed exeat system would assists the university administrator to utilize existing student information for managing and monitoring the students in the hall of residence through electronic administration and monitoring of exeat, the parents / guardian to be  mobile SMS  alerted  and prevention of  impersonation at the School  gate. The paper is organized into five sections: Section one Introduced the problem and justification of the proposed solution, Section two review related works in the problem domain, Section three provides materials method used to provide the solution, Section four describe how the method and materials were  implemented and tested while the Section five concludes the paper.

## 2.0 RELATED WORKS

A number of works has been done in the area of biometric technology and mobile communications over the years to the problem of entity entry/exit control. In Abdul Kadir et al (2009) proposed an RFID matrix card based auto identity system to the manual problem of monitoring student in boarding schools. Upon initial study of the three Boarding school in Malaysia, current process of maintaining students records in and out was not only tedious, misinformation always  happen as  students tend to provide inaccurate information. In Matjaz and Tusar (2007), a flexible modular system based on integration of arbitrary access sensors and an arbitrary number of stand-alone modules were applied to solve the problem of entity exit/entry. The system was tested with four sensors: a door sensor, an identity card reader, a fingerprint reader and a camera. However, identity cards can be lost, stolen and misused. Bochkov et al (2007) examined the security problem of identity theft where he specifically addressed the following issue: Why should we care about identity theft?; What options are available to solve this problem? What the solutions are?, and why some are more effective than others?. The authors discussed other biometric technologies that are emerging including vein patterns, facial thermographs, DNA typing, sweat pores, hand grip, fingernail bed, body odor, ear shape, gait pattern, skin luminescence, brain wave pattern, footprint recognition and foot dynamics. Also in relation to the identity theft, Vijay and Dattatray(2010) discussed the issue of using multimodal biometrics in systems. Biometric systems based on single source of information are called unimodal systems (i.e. using one source to access necessary information.

Of all the emerging biometric technologies, fingerprint identification is one of the most well-known and publicized biometrics because of their uniqueness and consistency over time, fingerprints have been used for identification for over a century, more recently becoming automated (i.e. a biometric) due to advancements in computing capabilities. It became popular as a means of identification and verification because of the inherent ease in acquisition, the numerous sources (ten fingers) available for collection, and their established use. According to Maltoni et al. (2003), fingerprint is one of the most mature biometric traits and considered legitimate proof of evidence in courts of law all over worldwide. Fingerprints are, therefore, used in forensic divisions worldwide for criminal investigations. More recently, an increasing number of civilian and commercial applications are either using or actively considering using fingerprint-based identification because of a better understanding of fingerprints as well as demonstrated matching performance than any other existing biometric technology. The discovery of uniqueness of fingerprints caused an immediate decline in the prevalent use of anthropometric methods of identification and led to the adoption of fingerprints as a more efficient method of identification(Lee et al., 1991).

With recent advances in internet and mobile technology, electronic service is becoming an important factor because different people can provide and obtain services without the limitation of location. The excellent e-service provides service via different channels and uses internet technology to provide customers with service in a cost effective manner. Customer communities are managed through e-mail, SMS messages, faxes etc. (Hua et al 2004). SMS Alert combines and integrates the benefits of cell phone technology and alarm and monitor systems, as it enables an individual to control and keep in contact with home, business or machinery and equipment. It instantly sends an SMS or missed call the moment the alarm is triggered. It can also be used to control and monitor machinery and equipment. Khiyal et al (2009) presented a method that focuses mainly on controlling home appliances remotely and providing security when the user is away from the place. The system is SMS based and uses wireless technology to revolutionize the standards of living and provides ideal solution to the problems faced by home owners in daily life.

An attempt has been made to review existing works on biometric implementation with a view of knowing the current tools used in its various application. Fingerprint technology is so far the most suitable and reliable approach for the system development as it basically takes care of security and prevents impersonation among students. It is less prone to forgery compared to the existing method on ground and hence can be deployed to solve the problem of student exeat at Bells University of Technology, Ota hall of residence.

## 3.0 MATERIALS AND METHODS
The following scientific approaches were used to achieve the central idea of this work. They are:
Requirement definition and infrastructural modeling

### 3.1 Requirement Definition of the Proposed Service Infrastructure
i. Mobile Students Exeat Monitoring and Management System Requirement
This requirement follows from the assumption that in order to automate the exeat management system, the system should provide: a) **Eligibility and Authentication:** The system should be designed in a way that only allows access to authorized personnel. b) **Uniqueness:** A student has only one exeat and it can not be used by another person. c) **Accuracy:** The administrator should be able to compute records and generate exeat reports with lesser errors. d) **Integrity:** students' exeat records can only be modified, updated or deleted by the assigned administrator. e) **Reliability:** The system should work robustly without any loss of records due to good and reliable database and also should be able to notify parents/guardians in lesser time. f) **Flexibility:** More modules expected of exeat operations can be integrated into the system to increase functionality. g) **Convenience**: students should be able to enter/exit the university with minimal sign in/out time.

**ii. Service Provision Requirement**
The infrastructure should allow the administrators to monitor registration of staff/supervisors and students, allow supervisors to grant exeat to students, record time of exeat granted, check student exeat number, monitor whether a particular student has returned after the exeat duration expired or not etc.

### 3.2 Infrastructural Model and Architect
i. **Overall System Architecture**
The Students Exeat Monitoring Automated Systems involves two important technologies namely: i) Biometric Fingerprint Technology (Scans the fingerprints of users) ii) SMS technology (automatically sends alert to parents or guardian). The hardware phase integrated into the system is the biometrics fingerprint scanner. The software phase is divided into two sub-phases: i) Front End (application interfaces the users would interact with) ii) Back End (database where the information is stored). In designing the front and back end of the system, some development tools required are: i) **Microsoft Visual Studio (.Net Framework):** The programming language used is C# which is an elegant and type-safe object-oriented language that enables developers to build a variety of secure and robust applications that run on the .NET Framework. C# can be used to create traditional Windows client applications, XML Web services, distributed components, client-server applications, database applications etc. ii) **Microsoft SQL Server 2005:** It is a fast, stable and true multi-user, multi-threaded SQL database server; SQL (Structured Query Language). This serves as the database at the back end because it is fast, robust and easy to use. Access to the database will be limited to the administrator in order to prevent unauthorized individual from having access to sensitive information. The designed system is a client-server system that describes a network in which processing is divided between a client program running on a user machine and a network server program. The system architecture of the designed system is shown in figure 1. The system components include fingerprint scanner, exeat management system, the system database and an SMS gateway.

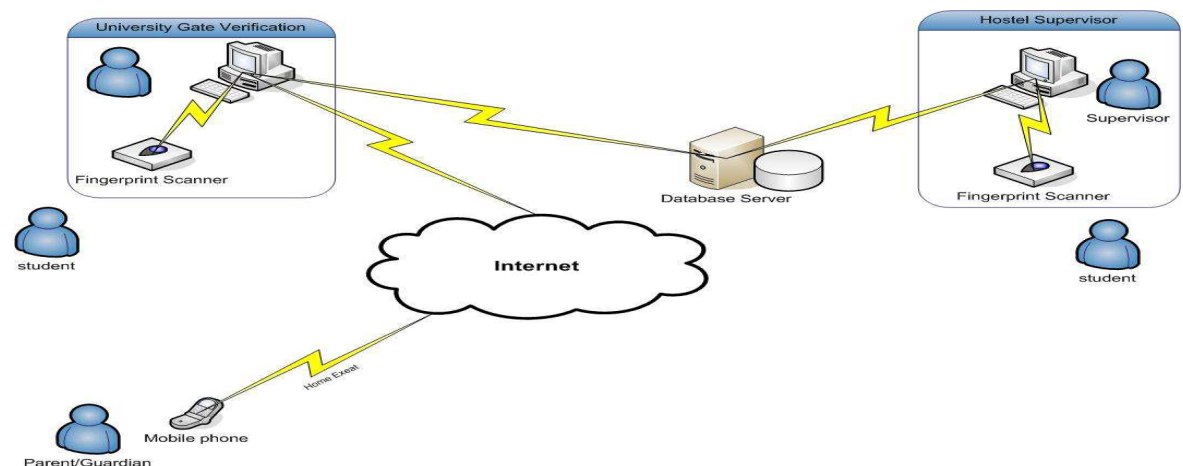

Figure 1: System Architecture of the Mobile Students Exeat Monitoring Systems

**Major System Components**

**The Fingerprint Scanner**
The Fingerprint scanner enrolls and verifies the identity of every person based on the marks on his or hers fingers and these marks have a pattern that cannot be changed or removed. The print is made up of ridges and furrows as well as characteristics that occur at minutiae points. Standard systems are comprised of a sensor for scanning a fingerprint and a processor which stores the fingerprint database and software which compares and matches the fingerprint to the predefined database. Within the database a fingerprint is usually matched to a reference number or Pin number which is then matched to a person's name. In instance of security, the match is generally used to allow or disallow access. (Thorton, 2000).

**Database Server:** In other to make comparison possible, the fingerprint representation and students matric number have to reside in a data repository. In this paper, a centralized database was used for storing each student's data. There will also be a link between the biometric data stored in the database to some information about the student's identity. When the database is queried, the feedback will not just include the biometric data, it also includes the personal information relating to the corresponding student and the database was implemented using Microsoft SQL Server 2005.

**SMS Gateway**
Message Alert format that is used is the Express Bulk SMS, with an SMS account opened. This will enable parents/ guardians get alerts when their ward takes home exeat. The exeat system is implemented as a server system. It enrolls, verifiers by granting exeat, and for the home exeat it sends an SMS over the internet to a number that has been specified in the database.

**ii. Biometric Authentication Framework**
Biometric authentication requires comparing a registered or enrolled biometric sample (biometric template or identifier) against a newly captured biometric sample. The biometric authentication system is used in two phases: The Enrolment phase and  The Verification phase.

**Enrolment Phase Design**
In the enrolment phase, a sample of the biometric trait is captured, processed by the computer and then stored in the system database for comparison at a later date. During this phase, the biometric characteristic of an individual is first scanned by a biometric reader to produce a raw digital representation of the characteristics. A quality checker is performed in to ensure that the required sample can be reliably compared during the verification stage. In order to facilitate matching, the raw digital representation is usually further processed by a feature extractor to generate a compact but expressive representation called a template. The template is then stored in the central database of the biometric system.

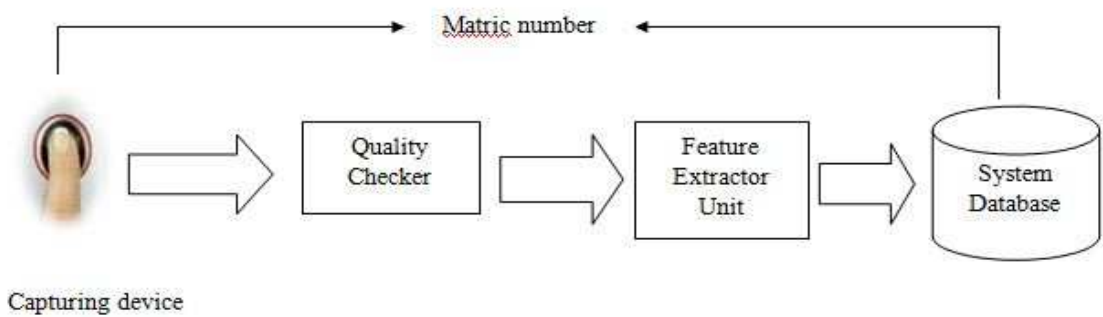

Figure 2: Enrolment Phase Design

**Verification Phase Design**
In the verification phase, the biometric system authenticates a student's claimed identity from their previously enrolled pattern and this is also called one-to-one matching. The verification task is responsible for verifying individuals at the point of access. During operation the students matric number, room number and the type of exeat (either day or home exeat) is inputted, at the gate the biometric reader captures the characteristics of the individual to be recognized and converts it to a digital format, which is further processed by feature extractor to produce a compact digital representation. The resulting representation is fed to the feature matcher, which compares it against the template of a single user retrieved from the system database.

In the biometric exeat system, the verification would be done at the gate after the supervisor has collected the student information and inputted it. At the gate a finger is used to authenticate the student and if there is match, the student is allowed to leave the school premises.

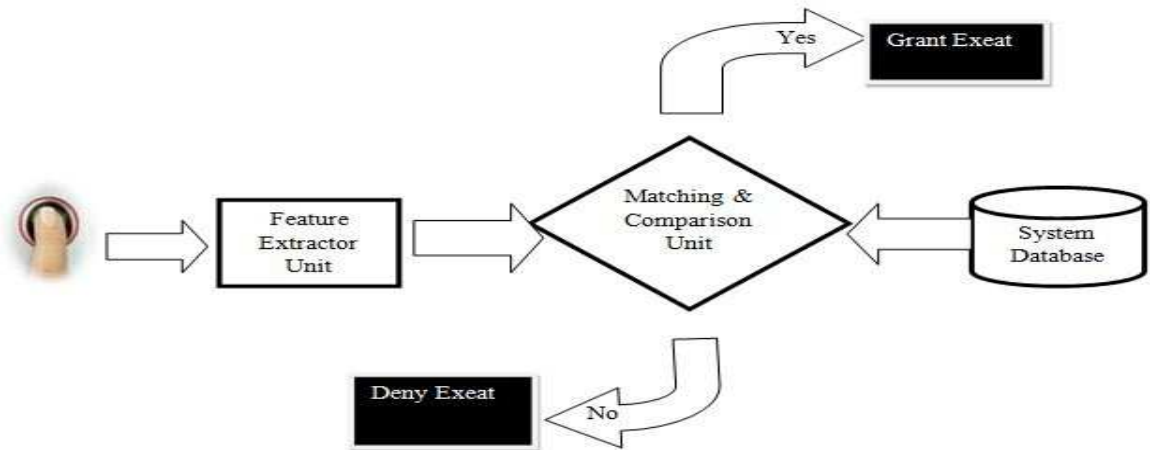

Figure 3: Verification Phase Design

### iii. Mathematical model for fingerprint identification

Fingerprint identification is based primarily on the minutiae, or the location and direction of the ridge endings and bifurcations (splits) along a ridge path. Fingerprints possess two main types of features that are used for automatic fingerprint identification and verification: (i) Ridge and furrow structure that forms a special pattern in the central region of the fingerprint and (ii) Minutiae details associated with the local ridge and furrow structure (Chander and Rajender,2005 ). Fingerprint images (FI) identification is achieved on the basis of template, containing the pattern features. Depending on different vendors, features of fingerprint image (FI) template are generally varied (Vladimir, 2007). Fingerprint Image template is processed in the form of

$$\Gamma : F_0^{(m)} \rightarrow \{L_m, L_l, L_r\}, \qquad (1)$$

Where $F_o^{(m)} = |f_0^{(m)}(x, y)|$ is the image skeleton.

$L_m$ is the minutiae list; $L_l$ is topological vectors list for lines; $L_r$ is ridge count vectors list for lines.

**Minutiae list**

Let $M_i$ – is minutiae which is indexed to number i . The minutiae list $L_m$ is in the following form

$$L_m = \{M_i = \{(x_i, y_i), \alpha_i, t_i, v_i, \theta_i, p_i, h_i\} | i \in 1..n_1\}, \qquad (2)$$

$|L_m| = n_1 -$ cardinal number; $(x_i, y_i), \alpha_i, t_i, v_i, \theta_i, p_i, h_i -$ coordinates, direction and type of minutiae as well as value and direction of curvature, probability and density of lines about minutiae.

**Topological vectors list**

Topological vectors list for lines is synthesized on the basis of all the nodes of skeleton, excluding minutiae nodes, and written in the form of

$$L_l = \{V_i = \{(e_j, n_j, l_j)\} | i \in 1..n_2, j \in 1..m_t\}, \qquad (3)$$

Where $V_i -$ topological vector for skeleton nodes cluster;

$|L_l| = n_2 -$ cardinal number and $n_2 > n_1$; $i -$ index like the number of topological vector; $j -$ number of link in topological vector; $e_j -$ event, and $l_j -$ length of link, formed with minutiae with number $n_j$; $m_t -$ quality of links taking into account central line in the form of

$$m_t = 4m + 2 \qquad (4)$$

**Ridge count vectors list**

Ridge count is calculated as the quantity of lines, placed on the straight line between two minutiae. In electron systems for one minutia Mi, as a rule, some similar values are determined.

$$L_r = \{R_i = \{(r_j, n_j)\} | i \in 1..n_3, j \in 1..n_4\}, \qquad (5)$$

where $R_i -$ is a vector of ridge count for the nodes group of the skeleton as ordered by index $j$ set of ordered pairs $(r_j, n_j)$; $|L_r| = n_3 -$ cardinal number and $n_3 > n_1$; $i -$ index as a number of vector and $n_4 < n_1$; $r_j -$ ridge count value, and $n_j -$ minutiae number on a link j.

iv. **Model Analysis**

The structure of the proposed system model can be analyzed using the use-case diagram, class diagrams sequence diagram and flowchart diagram. The use-case scenario of the infrastructure is shown in Figure 4 showing the interactions of the porters, guardians/parents and students on each tier of the model. The use case diagram has three actors. The registered porter login into the desired Mobile Students Exeat Monitoring and Management System Service, identified the student requesting exeat in the system, grant exeat to student. If the student desired to go home, an SMS alert is automatically sent to the guardian/parent to notify them. All information here is stored in the database and can be used for future reference. Figure 5 shows the Class diagram of the Mobile Students Exeat Monitoring System**,** Figure 6 shows the Sequence diagram of the Mobile Students Exeat Monitoring System and Figure 7 shows the Class diagram of the Students Exeat Monitoring System.

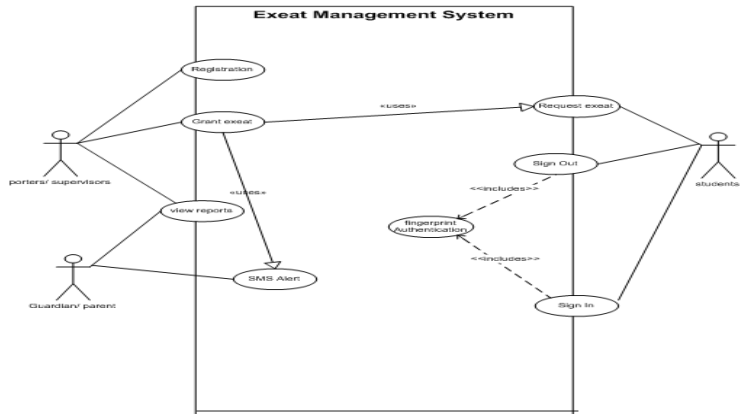

Figure 4: Use case diagram of the Mobile Students Exeat Monitoring Systems Using Fingerprint Biometric Authentication

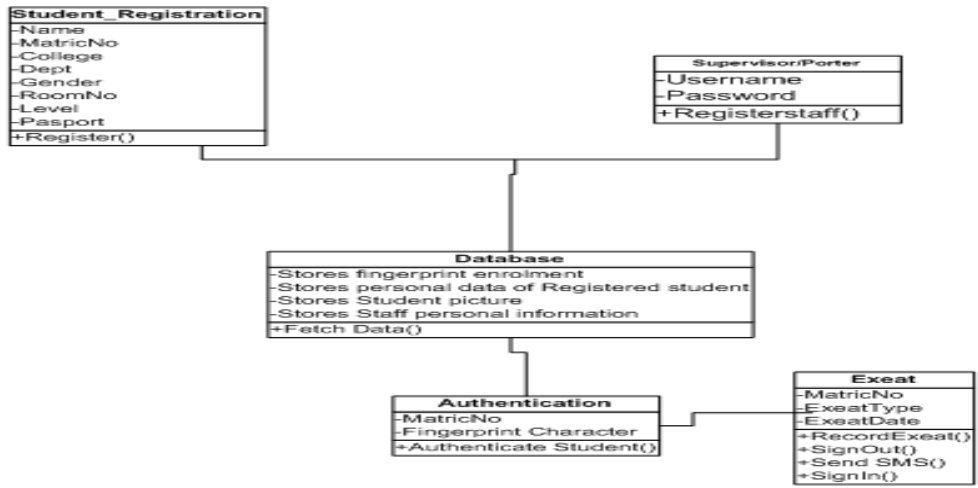

Figure 5: Class diagram of the Mobile Students Exeat Monitoring Systems Using Fingerprint Biometric Authentication

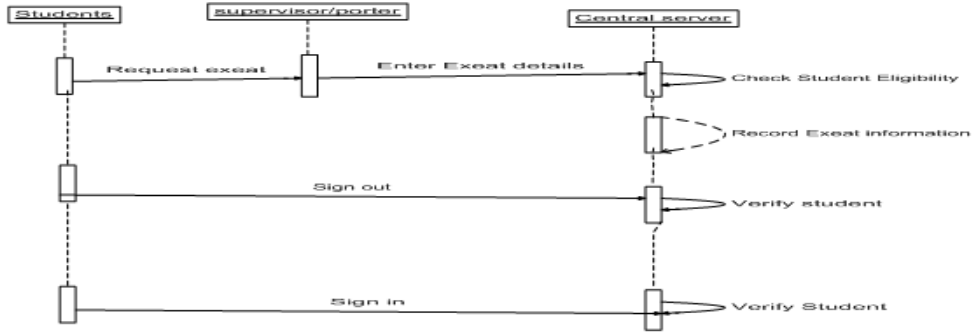

Figure 6: Sequence diagram of the Mobile Students Exeat Monitoring Systems Using Fingerprint Biometric Authentication

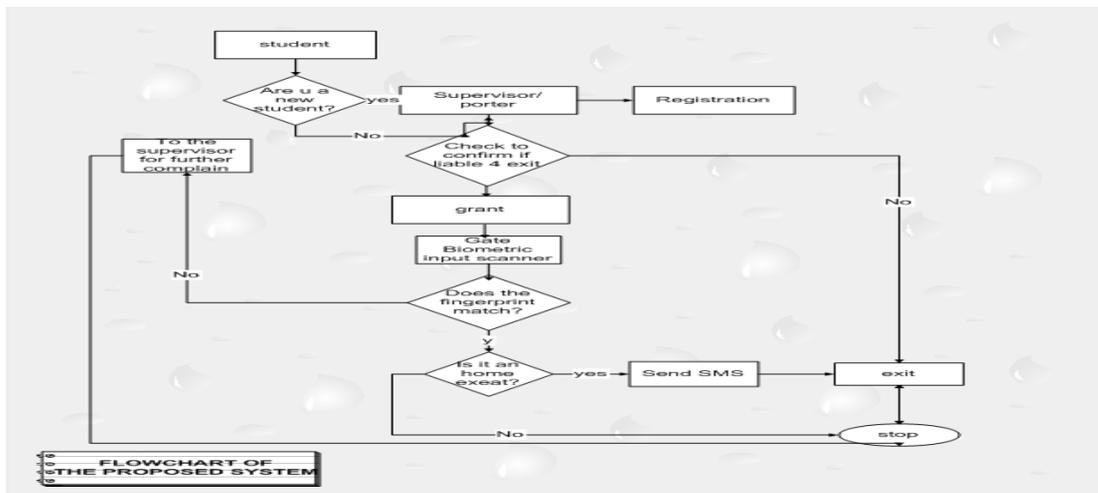

Figure 7: Class diagram of the Mobile Students Exeat Monitoring Systems Using Fingerprint Biometric Authentication

First draft
The Don International Journal of ICT and Youth Development (2012)  Vol 2 pp76 - 85

## 4.0 IMPLEMENTATION AND RESULT

The implemented system was tested using actual students' data in a typical university setting. During testing the staff logs in using his or user name and password and is allowed to grant day and home exeat, manage student records, enroll new students and then monitor if the student has returned at the expected time or not. The system grant the day exeat by verifying the fingerprint of the student and then it notifies the supervisor in charge the number of exeat that has been granted. For the home exeat it grants the exeat and then sends an SMS to the guardians phone number via the SMS gateway; stating that the student or ward has left school, and then when that student returns it verifies that the person has returned.

The exeat grant page is in two forms: the home and day exeat. This page is used to manage the type of exeat that the student has been granted. It keeps a record of the entire student and their information.  The student that requires the exeat is selected from the list and granted the exeat either day or home, when this is done the status of the student changes to: "student name" has not signed in.

For a successful request granted, in the case of the day exeat as shown in figure 8 the system displays the message "exeat granted" while for home exeat request it grants the system shows the message "now sending SMS" as shown in figure 9, once this is done, it shows the message "an SMS has been sent" to confirm the SMS delivery.

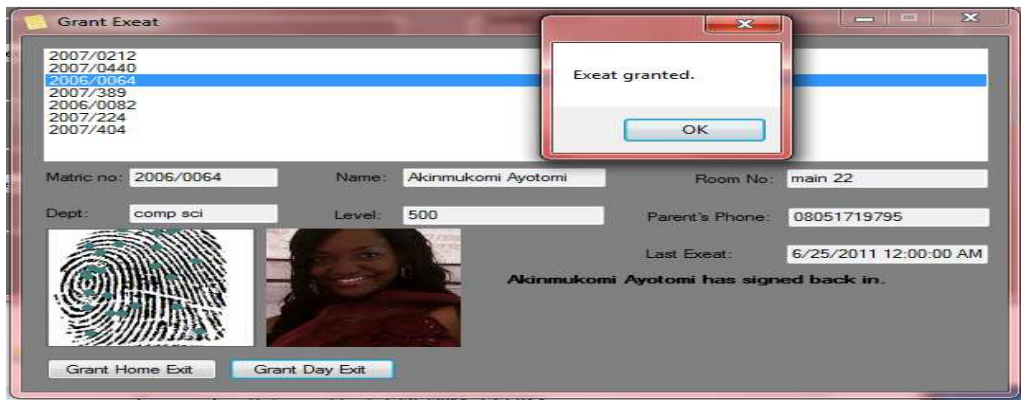

Figure 8: Granting day exeat

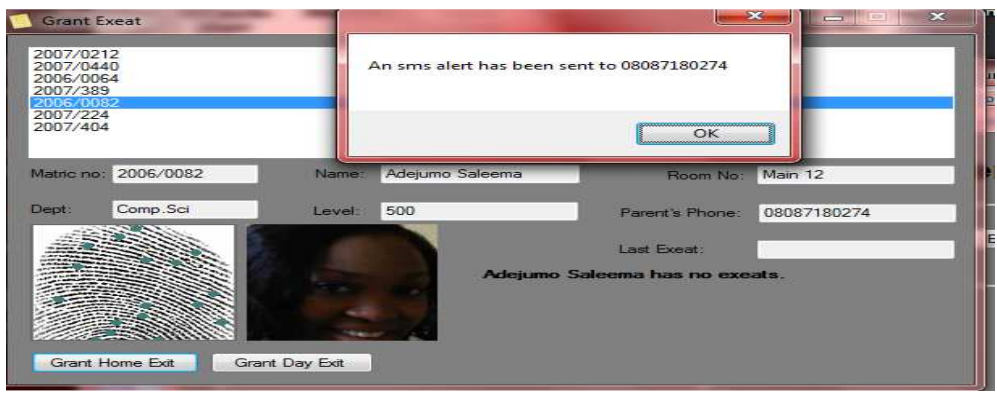

Figure 9: Granting home exeat.

A sample report page is shown in figure 10 and this page displays the student name, matric number, the type of exeat collected, expected date of return and finally if that student has returned or not. A returning student status is update once they confirm their return with their fingerprint thus, with this approach there is no problem of uncertainty as to whether a student returned or not as the system can automatically accurately monitor exeat.

Figure 10: Report page

## 5.0 CONCLUSION

In this paper, the proposed exeat based system was developed to solve inaccuracy, insecurity and impersonation challenges of the current paper based exeat system in use in most privately owned University Halls of Residence. With aid of a fingerprint biometric authentication, impersonation of other students is eradicated. The biometric device authenticates each student before granting access; ensuring that no student can take more the required number of the exeat per unit time. The mobile short message service informs the guardians the current location of where their wards are. The proposed system has improved the existing system in the following ways: Provide electronic solution to the existing method of issuing exeat to student which is paper based; discourage and prevent impersonation; Allow guardians or parents to monitor or be aware of their wards movement when they leave school; Eliminates the cost of making several copies of paper exeat and allow the university administrator to have a report of the student's exeat activities.

**REFERENCES**

Abdul Kadir H, Abdul Wahab M and Siti Nurul A (2009),"Boarding Students Monitoring Systems(E-ID) Using Radio Frequency Identification", Journal of Social Sciences, Volume 5(3),pp 206-211.

Bochkov. Y, Chiem. J, Sai. Y(2007). Use Biometric Techniques in Combating Identity Theft**.** Loyola Marymount University Los Angeles.

Chander Kant & Rajender Nath (2005). Reducing Process-Time for Fingerprint Identification System. International Journals of Biometric and Bioinformatics, Volume (3) : Issue (1)

Frischolz, R.and Dieckmann, U.(2000). A Multimodal Biometric Identification System ,IEEE Computer, Vol. 33, No. 2.

Hua, W,, and Lili, S.,Yanchun, Z., and Jinli, C. (2004) *Anonymous access scheme for electronic services.* In: 27th Australasian Computer Science Conference (ACSC2004), 18-22 Jan 2004, Dunedin, NZ.

Khiyal M.S, Aihab Khan, and Erum Shehzadi.(2009).Software Engineering Dept.,Fatima Jinnah Women University,Rawalpindi, Pakistan.

Lee, H.C, and Gaensslen, R,E. (1991). Advances in Fingerprint Technology, Elsevier, New York.

Maltoni, D, Maio, D, Jain, A.K|, Prabhakar, S.(2003). "Handbook of Fingerprint Recognition". Springer-Verlag,

Matjaž. G, Tušar. T (2007). Intelligent High-Security Access Control. Department of Intelligent Systems Jožef Stefan Institute.

Matyas.V., Rıha, Z. (2000). Biometric Authentication Systems. Technical report. Retrieved from http://www.ecom-monitor.com/papers/biometricsTR2000.pdf.

Omidiora E.O (2009) A prototype of an access control system for a computer laboratory 4$^{th}$ edition international conference on ICT to Teaching Research and Administration (AICTTRA 2009), ile-ife, Nigeria 4: (114-120).

The Global Information Technology Report (2009). World Economic Forum and INSEAD. ISBN 978-92-95044-19-7

Thornton, John (2000). Latent Fingerprints, Setting Standards In The Comparison and Identification. 84th Annual Training Conference of the California State Division of IAI. Retrieved from http://www.latent-prints.com/Thornton.htm.

Vijay M. M and Dattatray V. J (2007). Review of Multimodal Biometrics: Applications, challenges and Research Areas. International **Journal** of Biometrics and Bioinformatics (IJBB), Volume 3, Issue 5. 90.

Vladimir G..(2007). Mathematical Models of Fingerprint Image On the Basis of Lines Description. Department of Applied Mathematics  Chelyabinsk State University, Chelyabinsk, Russia.